\documentclass[twocolumn,preprintnumbers,amsmath,amssymb,floatfix,prd]{revtex4}
\usepackage{epsfig}
\usepackage{dcolumn}
\begin{document}

\title{Pion Fragmentation Functions at High Energy Colliders}
\author{Ignacio Borsa}
\email{iborsa@df.uba.ar} 
\author{Rodolfo Sassot}
\email{sassot@df.uba.ar} 
\affiliation{Departamento de F\'{\i}sica and IFIBA,  Facultad de Ciencias Exactas y Naturales, Universidad de Buenos Aires, Ciudad Universitaria, Pabell\'on\ 1 (1428) Buenos Aires, Argentina}

\author{Daniel de Florian}
\email{deflo@unsam.edu.ar} 
\affiliation{International Center for Advanced Studies (ICAS) and IFICI, UNSAM, 
Campus Miguelete, 25 de Mayo y Francia (1650) Buenos Aires, Argentina}

\author{Marco Stratmann}
\email{marco.stratmann@uni-tuebingen.de}
\affiliation{Institute for Theoretical Physics, University of T\"ubingen, Auf der Morgenstelle 
14, 72076 T\"ubingen, Germany}

\begin{abstract}
We revisit the description of pion production in proton-proton collisions
in the light of the very precise data taken at the Large Hadron Collider (LHC)
over the past decade. 
First attempts to include LHC results in next-to-leading order 
global QCD analyses of parton-to-pion fragmentation functions insinuated some 
conflict between data sets at different center-of-mass system energies.
We show that the data can be well described within their uncertainties by a consistent set of 
pion fragmentation functions once the theoretical scale dependence is taken into account 
in the global QCD analysis.
\end{abstract}
%
\maketitle

{\it Introduction.---} 
Pions are the most copiously produced particles in proton-proton collisions,
and their different parton content and distinctive decay modes make them invaluable 
tools to analyze the breakdown of the colliding protons and the subsequent 
recombination of their constituents into new hadrons \cite{Field:1976ve}. The precise 
measurement of these processes and their comparison with the most accurate predictions 
tests our understanding of hard scattering processes as described by the theory of strong interactions, 
Quantum Chromodynamics (QCD) \cite{Collins:1989gx}.

Fragmentation functions (FFs) play a cardinal role in the perturbative QCD description
of processes involving identified hadrons in the final state, as they connect 
the scattering of partons at short distance to the emerging hadrons observed in experiment \cite{Collins:1981uw}. 
From a phenomenological point of view, these functions parametrize the probabilities 
for the measured final state hadron to evolve from the partonic seed excited 
in the hard interaction \cite{Field:1976ve}. In the field theoretical framework, 
the nonperturbative, scale-dependent FFs are used to factorize any information on 
the hadronization process happening at long distances from the calculable partonic cross sections,
thereby canceling final state singularities at each order of perturbation theory \cite{Collins:1981uw}.
  
The notion of FFs is in complete analogy to that of the more familiar parton distribution functions (PDFs). 
In fact, the first attempts to infer them from data date back to the seventies, when combined PDFs and FFs extractions were first
explored \cite{Field:1976ve}. The experimental challenges to obtain precise data on hard 
processes with observed final state hadrons, not only well identified but also fully characterized
kinematically, hindered the progress in the determination of FFs compared to that of PDFs.
Until recently, FFs were obtained solely from the analysis of single inclusive  
electron-positron annihilation (SIA) data.

In 2007, a first global QCD analysis of FFs at next-to-leading order (NLO) accuracy
was presented in Ref.~\cite{deFlorian:2007aj}. It combined SIA data with pion production measurements 
performed in semi-inclusive deep inelastic scattering (SIDIS) and proton-proton collisions (PP) and
demonstrated the anticipated universality and factorization properties of FFs within the precision 
available at that time. Similar analyses have been published since then \cite{Sato:2016wqj,Bertone:2017tyb,Khalek:2021gxf}.

The analysis of Ref.~\cite{deFlorian:2007aj} was subsequently updated in \cite{deFlorian:2014xna} 
to account for the extremely precise SIA measurements from the Belle \cite{Belle:2013lfg} 
and BaBar \cite{BaBar:2013yrg} collaborations, SIDIS data from Compass \cite{COMPASS:2016xvm}, 
and the first PP results from the LHC  provided by the ALICE experiment \cite{ALICE:2012wos}. 

The results hinted at some degree of tension between the data sets, most noticeable between the 
LHC and the lower energy hadroproduction experiments at BNL-RHIC.
In fact, a slight tension was even observed between ALICE data taken at different 
center-of-mass system (c.m.s.) energies (0.9 TeV and 7 TeV, respectively), suggesting 
possible limitations in the assumptions made in the analysis, such as, for example, the accuracy 
of the NLO approximation employed, or even of the underlying factorization and universality 
property for FFs, the foundation of the theoretical description. 

In Ref.~\cite{deFlorian:2014xna} it was also shown that the tension between the experimental sets 
can be somehow alleviated by discarding PP data in the fit where the observed pion's
transverse momentum $p_T$ is less than about 5 GeV. In this region the NLO approximation is expected 
to be less and less reliable with decreasing $p_T$.
In addition, a very conservative normalization uncertainty, comparable in size to the large 
factorization scale dependence exhibited by the proton-proton cross sections, was introduced.
However, this strategy was unable to fully reconcile the $p_T$-dependence as predicted by the NLO 
approximation with the trend of the data. Since then, both ALICE and the RHIC experiments have delivered
additional, remarkably precise PP data on neutral and charged pion production at different 
c.m.s.\ energies. They not only confirmed their previous results but also clearly 
showed that the shortcomings of the FF analysis of Ref.~\cite{deFlorian:2014xna} 
further deepen with increasing c.m.s.\ energy.

In the following, we revisit pion production up to LHC energies in the light of the most recent 
measurements and by making use of up-to-date information on PDFs. 
We will show that the NLO framework can, contrary to the findings of Ref.~\cite{deFlorian:2014xna},
provide an accurate description of the world data not only at different c.m.s.\ energies but also
down to values of $p_T$ around 1 GeV. To this end, we need to exploit the
the factorization scale uncertainty that is inherent to any perturbative QCD estimate.
More specifically, we perform a NLO global analysis similar to those of 
Refs.~\cite{deFlorian:2007aj,deFlorian:2014xna}, that is based on the most recent sets of 
hard scattering data with identified pions in the final state and which explores different choices for the, in principle, 
arbitrary factorization scale at different c.m.s.\ energies. 
At variance with Refs.~\cite{deFlorian:2007aj} and \cite{deFlorian:2014xna},  
where the uncertainties of the obtained FFs were estimated using Lagrange multipliers and 
the improved Hessian technique, respectively, the present analysis implements a Monte Carlo sampling 
approach to produce a large sample of replicas for the FFs. 
In this way, estimates of the uncertainties inherited by any observable computed with our FFs 
can be obtained much more easily than in the Lagrange multipliers approach and without the complications 
related to choosing a particular tolerance criterion as in the Hessian framework.    

{\it Global analysis: setup and sets of data.---} 
As just described, crucial to our new analysis are the very precise PP data produced by ALICE 
at different c.m.s.\ energies $\sqrt{s}$, comprising $2.76,\,7,\,13\,\, \mbox{TeV}$ and 
$0.9,\,2.76,\,7,\,8\,\, \mbox{TeV}$ for charged and neutral pion production, respectively \cite{ALICE:2012wos,ALICE:2017nce,ALICE:2017ryd,ALICE:2020jsh}, as well as
the STAR and PHENIX data at $0.2\,\, \mbox{and}\,\, 0.51\,\, \mbox{TeV}$ 
\cite{STAR:2006xud,STAR:2009vxb,STAR:2011iap,STAR:2013zyt,PHENIX:2007kqm,PHENIX:2015fxo}. 
We also include the final $\pi^{\pm}$ SIDIS data by COMPASS instead of the preliminary sets 
still used in Ref.~\cite{deFlorian:2014xna}. 
In this way, our global analysis covers the wide range of c.m.s.\ energies spanned by the different 
PP experiments, the latest SIDIS data, as well as all the SIDIS and SIA data already available
in Ref.~\cite{deFlorian:2014xna}. 

The general strategy for our global analysis has been described in detail in 
Refs.~\cite{deFlorian:2007aj,deFlorian:2014xna} and need not be repeated here. 
It is based on the numerically efficient Mellin-moment technique that allows one to tabulate 
and store the computationally most demanding parts of a NLO calculation prior to the 
$\chi^2$-minimization from which we infer the optimum shapes of our FFs that suit best the data. 
The FFs for a parton of flavor $i$ into a positively charged pion 
are parameterized at our initial scale $Q_0=1\,\rm GeV$ as 
\begin{equation}
\label{eq:ff-input}
D_i^{\pi^+}\!(z,Q_0) =
N_i\, z^{\alpha_i}\, \sum_{j=1}^3 \gamma_j\, (1-z)^{\beta_{ij}}\;,
\end{equation}
where the free parameters $N_i$, $\alpha_i$, $\beta_{ij}$, and $\gamma_{ij}$ are 
determined by the fit. As usual, we assume charge conjugation symmetry, 
i.e., $D^{\pi^+}_q=D^{\pi^-}_{\overline{q}}$, and isospin symmetry for the unfavored FFs 
of the light sea quarks $D^{\pi^+}_{\overline{u}}=D^{\pi^-}_d$. 

The computation of precise partonic cross sections is a key ingredient in the quest 
for extracting FFs from data. One of the main issues in this respect is the need to
truncate the perturbative expansion at a given fixed order.
While next-to-NLO (NNLO) QCD corrections are available in the case of SIA, only NLO 
results are at our disposal for SIDIS and PP. For the latter, estimates for the size of 
the missing higher orders have been obtained by means of all-order expansions around 
the threshold for the partonic reaction \cite{Abele:2021nyo,deFlorian:2005yj}, 
finding rather large corrections along with a reduction of the scale dependence 
and, in case of PP, a much better agreement with the experimental data. 
But since the full set of NNLO corrections is still unavailable, we perform our analysis 
at NLO accuracy. For the same reason, we restrict ourselves to the 
massless quark approximation, even though heavy quarks mass effects have been shown to 
be sizeable in SIA analyses of pions and kaons \cite{Epele:2016gup,Epele:2018ewr}.

A possible way to get a rough estimate of the relevance of the missing higher order terms 
is by analyzing the dependence of the results on the unphysical renormalization 
and factorization scales for both the initial (PDFs) and final (FFs) state.
It is customary to set the default value of those scales to be of the same order as 
the typical physical scale ${\cal E}$ of the process, i.e., in our case, $\sqrt{s}$ for SIA, 
$Q$ for SIDIS, and $p_T$ for PP. Next, one varies the scales by a factor of 2 up and down; 
more sophisticated procedures can be found in \cite{Bonvini:2020xeo,Duhr:2021mfd}. 
In this paper, we follow the standard approach, choose all scales to be the same 
$\mu_{R}=\mu_{FI}=\mu_{FF}=\kappa\, {\cal E}$, and assume that 
any value of $\kappa$ between $1/2$ and $2$ is equally acceptable. 
On the one hand, this choice allows us to define a scale uncertainty band for each observable 
and, on the other hand, we can treat $\kappa$ as a free parameter for each experiment,
or group of experiments with similar kinematics, and let the fit select 
the values for $\kappa$ that yield an optimum reproduction of the data.
In this way, the long-standing issue with the description of the $p_T$-distributions 
in hadronic collisions \cite{dEnterria:2013sgr} can be solved. 

Finally, at variance with the analysis of Ref.~\cite{deFlorian:2014xna}, where the factorization 
scale dependence was included as a theoretical error in the $\chi^2$-minimization to which 
the Hessian method was applied, we determine first the optimum scale parameters $\kappa$ for each 
group of experiments and then produce for those scales a large set of replicas of the FFs by 
fitting them to statistically equivalent replicas of the data used in the fit. 
The optimum fit and its uncertainty estimates are assumed to be given by the statistical 
average of the obtained ensemble of replicas of the FFs and their corresponding variance, respectively,
thereby avoiding the arbitrariness of choosing a particular tolerance criterion.
This Monte Carlo sampling approach with a fixed functional form has been already successfully 
implemented for a global analysis of helicity parton densities in Refs.~\cite{DeFlorian:2019xxt,Borsa:2020lsz}. 

\begin{figure}[t]
\vspace*{-0.7cm}
\epsfig{figure=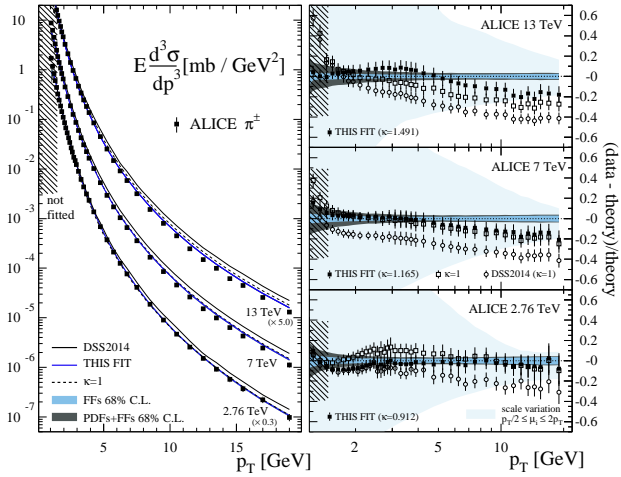,width=0.475\textwidth}
\vspace*{-0.5cm}
\caption{Left-hand side: Comparison of our best fit and other NLO results with ALICE $\pi^{\pm}$ PP data. 
Right-hand side: \textquotedblleft(data-theory)/theory" for each set of data and relevant sources of uncertainties 
for the new fit (shaded bands).
\label{fig:pp-pipm}}
\vspace*{-4mm}
\end{figure}

{\it Results.---} Figure~\ref{fig:pp-pipm} confronts the precise charged pion PP data 
from ALICE with our results at NLO computed with the newly obtained FFs and the
set of PDFs from Ref.~\cite{Bailey:2020ooq}. 
For comparison, we also show the outcome of a similiar fit with fixed $\kappa=1$ 
and a calculation based on the DSS set of FFs \cite{deFlorian:2014xna}. Both results 
clearly fail in reproducing the $p_T$-dependence of the data, with DSS
overshooting most of them by a large amount. The discrepancies are more noticeable 
with increasing c.m.s.\ energy.

\begin{figure}[b]
\vspace*{-0.7cm}
\epsfig{figure=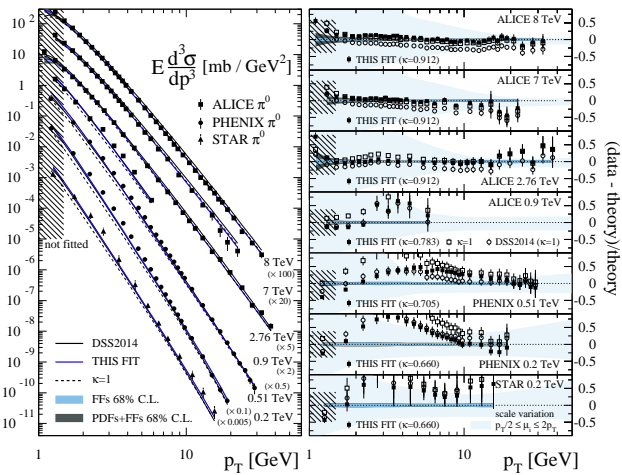,width=0.475\textwidth}
\vspace*{-0.5cm}
\caption{Same as in Fig.~\ref{fig:pp-pipm} but for $\pi^0$ production data from 
ALICE, PHENIX, and STAR. 
\label{fig:pp-pi0}}
\vspace*{-4mm}
\end{figure}

\begin{figure}[t]
\vspace*{-0.7cm}
\epsfig{figure=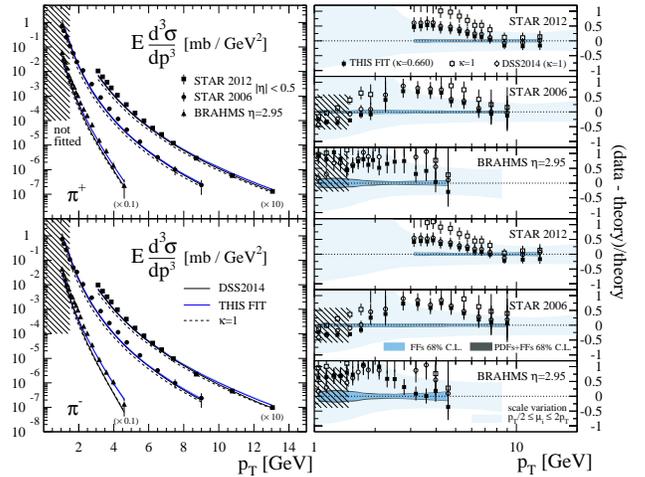,width=0.475\textwidth}
\vspace*{-0.5cm}
\caption{Same as Fig.~\ref{fig:pp-pipm} but for $\pi^{\pm}$ production data from STAR and BRAHMS
at different ranges of pseudorapidity $\eta$. 
\label{fig:pp-pip&m}}
\vspace*{-3mm}
\end{figure}

The fitted scale factors $\kappa$ that optimize the agreement with the data, 1.491, 1.165, and 
0.912 for $\sqrt{s}=13,\,7,$ and $2.76\,\mbox{TeV}$, respectively, are within the customary range 
and increase with increasing c.m.s.\ energy. The same values of $\kappa$ reproduce ALICE 
$\pi^0$ data as can be inferred from Fig.~\ref{fig:pp-pi0}. 
RHIC PP data on $\pi^0$ and $\pi^{\pm}$ yields, shown in Figs.~\ref{fig:pp-pi0} and \ref{fig:pp-pip&m}, 
respectively, prefer much smaller scale factors, $\kappa=0.660$ and 0.705 for $\sqrt{s}=0.2$
and $0.51\,\mbox{TeV}$, respectively, thus confirming the dependence on $\sqrt{s}$. 

The right hand sides of Figs.~\ref{fig:pp-pipm}-\ref{fig:pp-pip&m} provide for each set of data
a detailed \textquotedblleft(data-theory)/theory" comparison along with estimates
of the remaining relative uncertainties at NLO stemming from the FFs and, on top of the latter,
from the PDFs (blue and grey bands, respectively). Thanks to the new sets of data and the
lower cut in $p_T$ possible in our analysis, the FFs uncertainties of the present fit 
are significantly reduced compared to those found in Ref.~\cite{deFlorian:2014xna}.
The light blue bands indicate conservative estimates of the scale ambiguity 
relative to the $\kappa=1$ fit for a \textquotedblleft27 point", independent
variation of $\mu_{R}$, $\mu_{FI}$, and $\mu_{FF}$ within the standard range.
 
\begin{figure}[b]
\vspace*{-0.7cm}
\epsfig{figure=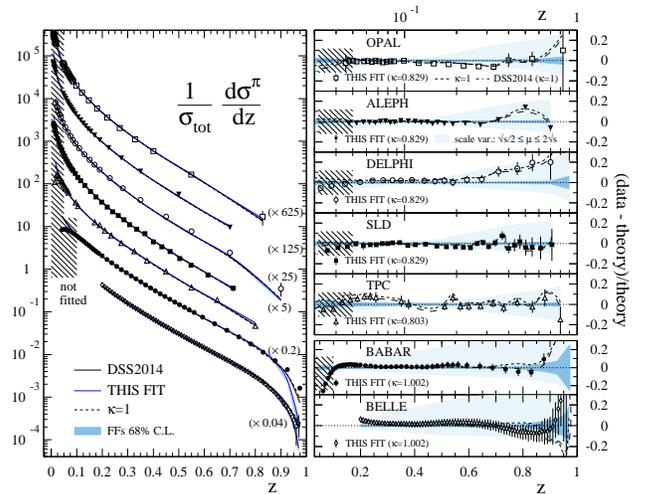,width=0.475\textwidth}
\vspace*{-0.5cm}
\caption{Similar as Fig.~\ref{fig:pp-pipm} but for SIA data from various experiments.   
\label{fig:sia}}
\vspace*{-4mm}
\end{figure}

\begin{figure}[h]
\vspace*{-0.4cm}
\hspace*{-0.4cm}
\epsfig{figure=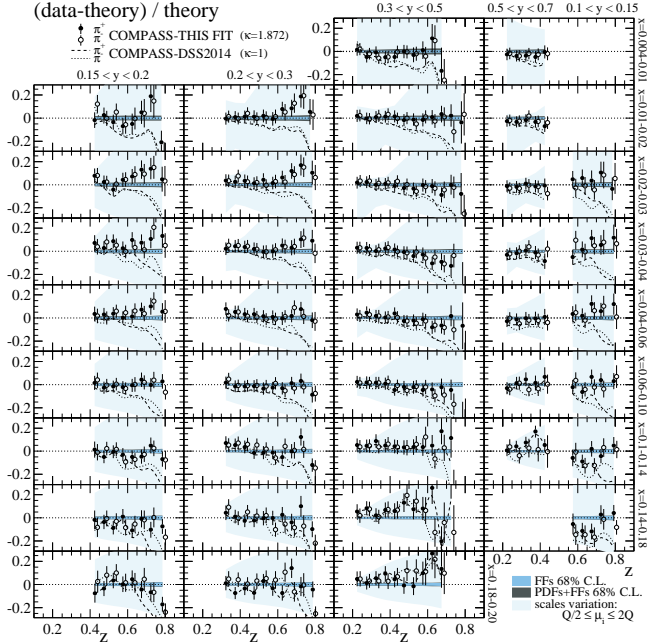,width=0.53\textwidth}
\vspace*{-1.0cm}
\caption{\textquotedblleft(data-theory)/theory" comparison of the DSS and our best fit to
the $\pi^{\pm}$ SIDIS multiplicities from COMPASS along with various uncertainty estimates (shaded bands).
\label{fig:sidis}}
\vspace*{-3mm}
\end{figure}

The much better agreement of the new FFs with the PP data compared to \cite{deFlorian:2014xna}
does not spoil the accord with neither SIA nor SIDIS data as can be seen in Fig.~\ref{fig:sia}
and \ref{fig:sidis}, respectively. In the fit of the SIA data we have again introduced scale
factors $\kappa$ (as indicated in the plot) for the different c.m.s.\ energies of the experiments 
in the range $\sqrt{s}=10.52$ GeV to $M_Z$. Note that all available fits describe the 
SIA data equally well.

\begin{figure}[b]
\vspace*{-0.3cm}
\hspace*{-0.3cm}
\epsfig{figure=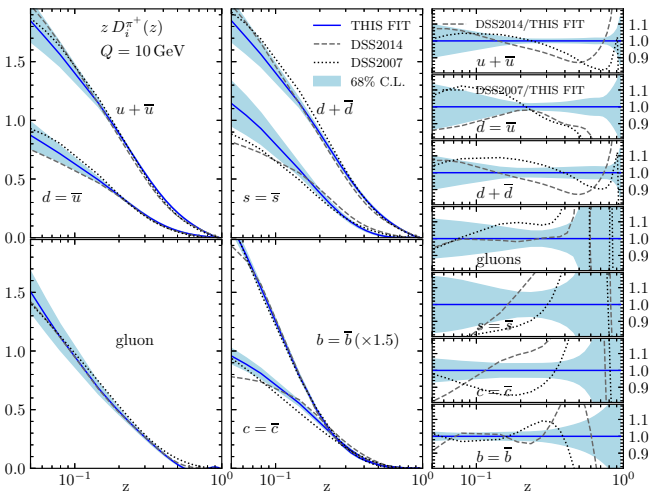,width=0.50\textwidth}
\vspace*{-0.8cm}
\caption{The obtained individual FFs for $\pi^+$ at scale $Q=10\,\mathrm{GeV}$ along with 
uncertainty estimates at 68\% C.L.\ compared to previous analyses by DSS \cite{deFlorian:2007aj,deFlorian:2014xna}.
The panels on the right-hand-side show the corresponding relative uncertainties. 
\label{fig:dist}}
\vspace*{-3mm}
\end{figure}

\begin{table}[h]
\caption{\label{tab:exppiontab} Data sets, normalizations $N_i$ as defined in Eq.~(6) of \cite{deFlorian:2014xna}, 
and the partial and total $\chi^2$ values obtained in the fit. }
\begin{ruledtabular}
\begin{tabular}{lrrrr}
experiment& data & $N_i$  & \#data & $\chi^2$ \\
          & type &        & in fit     &         \\\hline
{\sc Tpc} \cite{ref:tpcdata}\hfill 29~GeV  & incl.\            &  1.038 & 17 & 22.0 \\          
              \multicolumn{2}{r}{ $uds,c,b$ tag}         &  1.038 &  27 & 16.6\\
{\sc Tasso} \cite{ref:tassodata}\hfill 34~GeV &   incl.    & 1.038     & 11 &  28.4    \\         
                                \hfill 44~GeV &   incl.    & 1.038     &  7 &  20.8    \\ 
{\sc Sld} \cite{ref:slddata}\hfill 91.2~GeV  & incl.\  &  0.977 & 28 & 19.5 \\   
           \multicolumn{2}{r}{  $uds,c,b$ tag}         &  0.977 & 51 & 39.2 \\
{\sc Aleph} \cite{ref:alephdata}\hfill 91.2~GeV    & incl.\  & 1.012 & 22 & 44.5\\   
{\sc Delphi} \cite{ref:delphidata}\hfill 91.2~GeV   & incl.\  & 1.000  & 17 & 23.0\\  
                      \multicolumn{2}{r}{ $uds,b$ tag}   &  1.000  & 34 & 33.8 \\
{\sc Opal} \cite{ref:opaldata}\hfill 91.2~GeV   & incl.\ & 1.000 & 21 & 31.5 \\   
 \multicolumn{2}{r}{$u,d,s,c,b$ tag} &  0.793  & 25 & 62.1 \\
{\sc BaBar} \cite{BaBar:2013yrg}\hfill 10.54~GeV    & incl.\ &   1.060 & 45  & 142.4 \\ 
{\sc Belle} \cite{Belle:2013lfg}\hfill 10.52~GeV      & incl.\ &   1.067 & 78  & 60.2 \\    \hline 
\textbf{SIA data (sum)}                  &        &         & 378 &  544.1       \\    \hline    
{\sc Hermes} \cite{HERMES:2012uyd}\hfill  $\pi^+$,$\pi^-$ &(p-$Q^2$)&  0.984 & 56& 54.3\\
                                 \hfill   $\pi^+$,$\pi^-$ &(d-$Q^2$)&  0.988 & 56& 46.5\\
                                 \hfill  $\pi^+$,$\pi^-$  &(p-$x$)  &  1.007 & 56& 159.5\\
                                 \hfill  $\pi^+$,$\pi^-$ & (d-$x$)  &  1.009 & 56& 189.5 \\
{\sc Compass} \cite{COMPASS:2016xvm}  \hfill $\pi^+$,$\pi^-$ &(d-z) &  1.004 & 510 &  302.1    \\ \hline    
\textbf{SIDIS data (sum)}                    &        &         & 734 &    751.9      \\    \hline    

{\sc Brahms} \cite{BRAHMS:2007tyt} \hfill 0.20~TeV  & $\pi^+$,$\pi^-$            & 1.313 &  26  & 13.5       \\    
{\sc Star} \cite{STAR:2006xud,STAR:2009vxb,STAR:2011iap,STAR:2013zyt} 
    \hfill 0.20~TeV                 & $\pi^{0}$                        & 1.190 &  12  & 8.2       \\    
     \hfill 0.20~TeV           & $\pi^{0}$                        & 0.921 &  7  & 4.0        \\    
    \hfill 0.20~TeV                   & $\pi^+$,$\pi^-$                  & 1.029 &  26  & 37.3        \\    
    \hfill 0.20~TeV                & $\pi^+$,$\pi^-$                  & 1.158 &  34  & 73.8       \\    

{\sc Phenix} \cite{PHENIX:2007kqm,PHENIX:2015fxo}
                                  \hfill 0.20~TeV    & $\pi^0$ &  1.177 & 22 &  13.8  \\
                                  \hfill 0.51~TeV    & $\pi^0$ &  1.178 & 27 &  32.9 \\

{\sc Alice} \cite{ALICE:2012wos,ALICE:2017nce,ALICE:2017ryd,ALICE:2020jsh}
                                 \hfill 0.90~TeV & $\pi^0$          & 1.012 &   7 & 52.0        \\ 
                                 \hfill 2.76~TeV& $\pi^0$           & 1.002 &  24 & 17.4        \\ 
                                 \hfill 2.76~TeV& $\pi^{\pm}$       & 0.959 &  38 & 15.6        \\ 
                                 \hfill 7~TeV   & $\pi^0$           & 1.016 &  25 & 30.6        \\ 
                                 \hfill 7~TeV   & $\pi^{\pm}$       & 0.976 &  32 & 23.9        \\ 
                                 \hfill 8~TeV   & $\pi^0$,          & 1.048 &  36 & 34.5       \\ 
                                 \hfill 13~TeV  & $\pi^{\pm}$       & 0.981 &  32 & 56.2        \\  \hline
\textbf{{PP data (sum)}}                      &        &         & 348 & 413.7        \\    \hline

\hline\hline
\textbf{TOTAL:} & & & 1460 & 1709.7\\
\end{tabular}
\end{ruledtabular}
\end{table}

The present fit includes the final $\pi^+$ and $\pi^-$ SIDIS multiplicities from COMPASS, 
which are superior both in number and precision to the preliminary set used in \cite{deFlorian:2014xna}.
Our analysis includes data for $Q^2 \ge 1.5\,\mathrm{GeV}^2$ and neglects deuteron nuclear 
corrections \cite{Epele:1991np} throughout. The quality of the fit is illustrated
in Fig.~\ref{fig:sidis} in terms of a \textquotedblleft(data-theory)/theory" comparison for each
kinematic bin along with estimates for the relevant relative uncertainties of the fit (shaded bands).
The data prefer a scale factor $\kappa=1.872$, while the corresponding Hermes data \cite{HERMES:2012uyd} 
(not shown \cite{upon-request}) are best reproduced with $\kappa=1.402$.
 
Figure~\ref{fig:dist} compares the shape of the newly obtained FFs for the different flavors 
to previous extractions by DSS and shows their absolute and relative uncertainties (shaded bands).
None of the distributions is dramatically different than previous extractions, 
except in regions where they are still poorly constrained by data, but the remaining
uncertainties are significantly reduced, especially for the gluon FF.

Finally, Table~\ref{tab:exppiontab} summarizes the data sets used in our NLO global analysis, 
the computed normalization shifts $N_i$ as defined in Eq.~(6) of Ref.~\cite{deFlorian:2014xna}, 
and the $\chi^2$-values. 

{\it Discussion and conclusions.---} We have shown that the same set of nonperturbative 
pion FFs that reproduces the world data on SIA, SIDIS, and PP at RHIC kinematics 
can also fully account for the recent, very precise measurements of pion production 
cross sections up to highest energies available at the LHC.
The obtained constraints on the FFs from this large set of data are significant.

The key ingredient of our new global analysis at NLO accuracy of QCD is to fully exploit
the theoretical scale dependence, that determines how much of a measured cross section
is attributed to nonperturbative quantities such as FFs.
This choice is in principle arbitrary when applied in a consistent fashion.
Since the inclusion of different data sets taken at various energy scales in global fit
makes the choice of scale not obvious, we let the data determine, within the conventional range,
what optimum value is preferred for each set of data.
The result is a new fit of parton-to-pion FFs with increased precision and significantly reduced uncertainties
that turns one-particle-inclusive processes and FFs to a much better tool to unveil 
new aspects of hadron structure in the next generation of experiments such as the Electron-Ion Collider.

%
I.B.\ wishes to thank 
the University of Tuebingen for hospitality during the final stages of the work.
This work was supported in part by CONICET, ANPCyT, UBACyT, and
by Deutsche Forschungsgemeinschaft (DFG) 
through the Research Unit FOR 2926 (Project No. 40824754). 
%


\begin{thebibliography}{99}
\bibitem{upon-request}
Further results and the set of replicas of our FFs are available upon request from the authors.
\end{thebibliography}
\end{document}